\documentclass[reprint,amsmath,amssymb,aip,jcp,
twocolumn, floatfix]{revtex4-2}
\usepackage[normalem]{ulem}
\usepackage[dvipsnames]{xcolor}
\usepackage[section]{placeins}
\usepackage{graphicx} 
\usepackage{siunitx}
\usepackage{derivative} 
\usepackage{mathtools}
\usepackage{amssymb}
\usepackage{bm}
\usepackage{lmodern}
\usepackage{booktabs}
\usepackage[utf8]{inputenc}
\usepackage{adjustbox}
\usepackage{anyfontsize}
\usepackage{ulem}
\usepackage[english]{babel}
\usepackage{subcaption}
\usepackage[protrusion=true,
            expansion=true,
            final,
            babel]{microtype}
\definecolor{linkcolor}{rgb}{0.0,0.3,0.5}
\usepackage[
unicode, 
colorlinks=true,
linkcolor=linkcolor,
citecolor=linkcolor,
filecolor=linkcolor,
urlcolor=linkcolor,
hyperfootnotes=true]{hyperref}
\usepackage{cleveref}

\newcommand{\kT}{k_\text{B}T}
\newcommand{\rv}{{\mathbf r}}

\newcommand{\rhob}{\rho_\mathrm{b}}
\newcommand{\rmex}{{\mathrm{ex}}}

\newcommand{\FE}{\mathcal{F}}
\newcommand{\req}{\rho_\mathrm{eq}}
\newcommand{\reqs}{\rho_\mathrm{MC}}
\newcommand{\hs}{\mathrm{hs}}
\newcommand{\ex}{\mathrm{ex}}

\newcommand{\bond}{\mathrm{bond}}

\newcommand{\np}{N_\mathrm{p}}
\newcommand{\tj}[6]{
\begin{pmatrix}
  #1 & #2 & #3 \\
  #4 & #5 & #6 
\end{pmatrix}}

\newcommand*\conj[1]{\ensuremath{\overline{#1}}}

\newcommand{\nav}{n}
\newcommand{\ntwo}{{\mathbf n}_2}
\newcommand{\wtwo}{{\mathbf w}_2}

\newcommand{\be}{\begin{equation}}
\newcommand{\ee}{\end{equation}}
\newcommand{\bea}{\begin{eqnarray}}
\newcommand{\eea}{\end{eqnarray}}
\newcommand{\bml}{\begin{multline}}
\newcommand{\eml}{\end{multline}}

\begin{document}
\title{The orientational structure of a model patchy particle fluid: simulations, integral equations, density functional theory and machine learning}
\author{Alessandro Simon}
\email{alessandro-rodolfo.simon@uni-tuebingen.de}
\affiliation{Institute for Applied Physics, University of Tübingen, Auf der Morgenstelle 10, 72076
Tübingen, Germany}
\author{Luc Belloni}
\affiliation{LIONS, NIMBE, CEA, CNRS, Universit\'e Paris-Saclay, 91191 Gif-sur-Yvette,
France}
\author{Daniel Borgis}
\affiliation{Maison de la Simulation, USR 3441 CNRS-CEA-Universit\'e Paris-Saclay,
91191 Gif-sur-Yvette, France}
\affiliation{PASTEUR, D\'epartement de chimie, \'Ecole normale sup\'erieure, PSL University,
Sorbonne Universit\'e, CNRS, 75005 Paris, France}
\author{Martin Oettel}
\email{martin.oettel@uni-tuebingen.de}
\affiliation{Institute for Applied Physics, University of Tübingen, Auf der Morgenstelle 10, 72076
Tübingen, Germany}
\begin{abstract}
We investigate the orientational properties of a homogeneous and inhomogeneous tetrahedral 4-patch fluid (Bol--Kern--Frenkel model). Using integral equations, either (i) HNC or (ii) a modified HNC scheme with simulation input, the full orientational dependence of pair and direct correlation functions is determined. Density functionals for the inhomogeneous problem are constructed via two different methods. The first, molecular density functional theory, utilizes the full direct correlation function and an isotropic hard-sphere bridge functional. The second method, a machine learning approach, uses a decomposition of the functional into an isotropic reference part and a mean-field orientational part, where both parts are improved by machine learning techniques. Comparison to simulation data at hard walls and around hard tracers show a similar performance of the two functionals. 
Machine learning strategies are discussed to eliminate residual differences, with the goal of obtaining machine-learning enhanced functionals for the general anisotropic fluid.  
\end{abstract}
\maketitle
\tableofcontents
\section{Introduction}
Classical fluids made of particles with strong directional bonding (associating fluids) are an important class of systems relevant for the understanding of polymerization, self-assembly in molecular and colloidal systems and phase behavior of proteins. A simple model system for associating fluids are spherical patchy particles made of hard spheres with $\np$ localized spots (patches) on their surface, which lead to strong attraction when they overlap. 

Specifically, patchy particles with $\np=4$ patches are interesting model systems for tetrahedrally coordinating particles and thus the system has also some relevance to molecular fluids such as water \cite{Russo2022}. Patchy particles are becoming increasingly popular as colloidal models to explain certain features of protein solutions. An example is the phase and adsorption behavior of {bovine serum albumine (BSA)} in solutions with trivalent cations \cite{Roosen2014,Fries2017,Fries2020}. In these systems, the proteins bind at specific sites with the cations acting as a bridge, and can be described as ion-activated patchy particles.   Regarding bulk thermodynamic behaviour we note that  
the equation of state and consequently the phase diagram of these patchy particles can be understood very well using a statistical theory of bonding developed by {Wertheim, namely first order thermodynamic perturbation theory  (TPT1),} which only needs the pair correlations in a reference hard sphere system \cite{wertheim1987thermodynamic}. Gas--liquid binodals from TPT1 and simulations agree on a semiquantitative level \cite{Bianchi2006,Bianchi2008,Romano2010}. The dynamic properties of the fluid phase (relaxation) are strongly different from the ones in simple fluids, due to the formation of transient networks and have been mainly studied by simulations addressing the time-dependent bonding statistics \cite{Roldan2017}.    

Although patchy particles (if besides the attractive spots no other isotropic attractions are assumed) are in some sense ``maximally anisotropic'' in their attractions, the intrinsic orientational order in the equilibrium fluid has not been investigated very much in the past. As described, Wertheim's TPT1 does not need it and simulations focus on the bonding statistics which is easier to access but is of course a consequence of the orientational order. Explicit account for orientational structure appears to be necessary when discussing solvation. 
Solvated objects such as solute particles or surfaces~\cite{stopper2018bulk} may impose very strong directional constraints.    

Theoretical methods for the description of the orientational structure in anisotropic fluids can be roughly classified into integral equations (IE) and classical density functional theory (DFT) approaches. The theoretical background for IE's has been worked out in the 1970's \cite{gray84theory,blum1972invariant}, mostly with the aim of describing molecular fluids. Applications to realistic associated fluids such as water dates from the late 1990s~\cite{Richardi1999,Lombardero1999}. Such applications to patchy fluids have not been undertaken to the best of our knowledge. For patchy particles, an alternative IE approach is the ``multidensity'' formalism which considers orientationally averaged correlations but resolves them with regard to the binding state (i.e. number of bonds) of the particle, see Ref.~\cite{Kalyuzhnyi2020} and references therein.      
Classical DFT methods for patchy particles specifically have focused on functionals of the orientationally averaged density, taking into account the orientational association behavior in an averaged manner, based on Wertheim’s association theory~\cite{yu2002fundamental, stopper2018bulk}. Recent work~\cite{simon2024machine} has shown that at hard walls, a 4-patch Kern--Frenkel fluid shows strong orientational correlations close to hard walls, which cannot be captured by orientationally averaged DFT, for a similar finding see also Ref.~\cite{Gnan2012} or Ref.~\cite{teixeira2019patchy} for the case of two patches. Therefore, resolving the orientational dependence in the density profile appears to be necessary to correctly account for solvation properties of patchy particles. RPA-like functionals (which are second-order in the density) do not work well for patchy particles \cite{simon2024machine}. An alternative is to use Taylor-expanded functionals around a reference bulk fluid, which in their second-order term need the full orientation-resolved direct ({pair}) correlation function as input (which must be provided by an IE approach). Higher-order terms (``bridge functional'') may be approximated by the corresponding bridge functional of an isotropic system. In the past years, this approach has been developed and applied successfully for aqueous systems (``molecular DFT'' \cite{Zhao2011,Ding2017,Borgis2020,Borgis2021}) but the methodology can also be transferred to patchy particles.

In this work, the orientational correlations of a model 4-patch fluid are investigated for the bulk and close to a hard wall. For the bulk pair correlations and the density profile at a hard wall, Monte Carlo simulation data are obtained as a reference. Direct correlation functions (DCFs) cannot be simulated directly but are obtained via an IE scheme developed in Ref.~\cite{belloni2017exact}. These DCFs are input for the molecular DFT approach which is then applied to the solvation problem of hard walls and hard solutes. Results are compared to an improved approach along the lines of Ref.~\cite{simon2024machine}, where the functional is split into an orientationally averaged and a mean-field part taking into account the orientational correlations. In Ref.~\cite{simon2024machine}, the kernel of the mean field part was determined using machine learning (ML) methods and the reference part was the fundamental measure theory (FMT)--based ``Stopper--Wu'' functional of Ref.~\cite{stopper2018bulk}. Here, the FMT weight functions in the reference functional are improved by ML methods. Overall, we find rather good agreement of both DFT's with the hard wall simulation data, with ``molecular DFT'' performing better (and in excellent agreement with simulations) for the orientational correlations, and ML-improved ``Stopper--Wu'' DFT performing slightly better for the orientationally averaged density profile. The comparison between the two DFT's offers also some insight into the orientational expansion of the direct correlation function.

The paper is structured as follows: Sec.~\ref{sec:model} defines the model (a symmetric 4-patch Kern--Frenkel model), introduces orientational expansions for density profile and pair correlation functions tailored to the symmetries of the model and briefly describes the simulations. In Sec.~\ref{sec:ie}, the IE methods are introduced and results for pair correlation and bridge functions in the supercritical region are discussed. Sec.~\ref{sec:dft}  discusses two approaches for the construction of an orientation–resolved density functional for the general inhomogeneous problem. The first (molecular DFT), is based on a functional Taylor expansion and requires as input the fully orientation-resolved direct correlation function. 
The second approach employs a machine-learned reference functional depending on the angular–averaged density and an orientation-dependent remainder parameterized in mean–field form. The density functionals are applied to hard wall inhomogeneities. Finally, Sec.~\ref{sec:summary} summarizes and concludes the work.

%The inhomogeneous pair
%distribution function that is necessary for the approach are 

\section{Model definitions and simulations}
\label{sec:model}

\subsection{Bol--Kern--Frenkel model}

The Bol--Kern--Frenkel \cite{bol1982monte, kern_frenkel2003} (BKF) potential, often just referred to as the Kern--Frenkel (KF) potential $v$ between a pair of particles with radius $R$ each is a sum of a hard sphere part $v_\mathrm{HS}(r)$ for diameter $\sigma=2R$ and
an anisotropic patch part $v^\mathrm{KF}$ that depends on the individual orientation of the particles and their
positions relative to each other. For parametrizing the orientation of a three dimensional body
we will use Euler angles with the angles
$(\phi, \theta, \chi) \equiv \Omega$ and 
\begin{equation}
 \int \odif{\Omega} \equiv \int_0^{2\pi} \odif{\phi} \int_0^\pi \odif{\theta} \int_0^{2\pi} \odif{\chi} = 8 \pi^{2} \equiv K.    
 \label{eq:int_omega}
\end{equation}

\begin{figure}
    \centering
    \includegraphics[width=.7\linewidth]{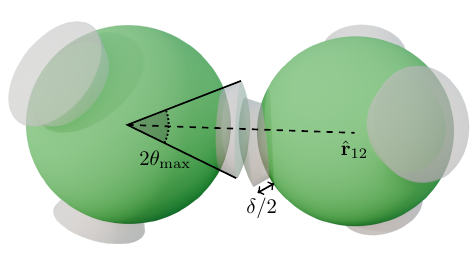}
    \caption{Patch--patch interaction in the Kern--Frenkel model between two particles with center-to-center vector $\hat{\rv}_{12}$. The angle $\theta^\mathrm{max}$ specifies the opening of the patch cone and the parameter
     $\delta$ the extension in the radial direction.}
    \label{fig:kf}
\end{figure}

The anisotropic part of the potential is defined by
\begin{equation}
  \label{eq:16}
  v^\mathrm{KF}(\rv_{12}, \Omega_{1}, \Omega_{2}) =  \phi_{\mathrm{sw}}(r_{12})\!\!\sum_{\alpha,\beta=1}^{N_\mathrm{p}}\!\!  \phi_{\mathrm{p}}(\rv_{12}, \hat{\rv}_{1}^{\alpha}(\Omega_{1}), \hat{\rv}_{2}^{\beta}(\Omega_{2}))\,.
\end{equation}
Here, $\np=4$ is the number of patches and   $\phi_{\mathrm{sw}}(r)$ is a square well potential
\begin{equation}
%  \phi_{\mathrm{sw}}(r)~=~ \left\{ \begin{matrix} -\epsilon \qquad (r \in [\sigma,\sigma+\delta]) 
   %    \\ 0 \qquad \mathrm{(otherwise)} \end{matrix} \right. \,, \label{eq:sw_patch}
  \phi_{\mathrm{sw}}(r) =
  \begin{cases}
    -\epsilon &\text{if } r \in [\sigma,\sigma+\delta] \\
    \phantom{-}0&\text{otherwise}
  \end{cases}
  \label{eq:sw_patch}
\end{equation}
with the attraction range $\delta$ and attraction depth $\epsilon$. $\phi_{\mathrm{p}}$ is the orientational patch--patch part
\begin{multline}
  \label{eq:17} %
  \phi_{\mathrm{p}}(\rv_{12}, \hat{\rv}_{1}^{\alpha}(\Omega_{1}), \hat{\rv}_{2}^{\beta}(\Omega_{2}))  = \\\begin{cases}
    1 &\text{if both}\begin{cases}
      \phantom{-}\hat{\rv}_{12} \cdot \hat{\rv}_{1}^{\alpha} > \cos \theta_{\mathrm{max}}\\
      -\hat{\rv}_{12} \cdot \hat{\rv}_{2}^{\beta} > \cos \theta_{\mathrm{max}}
    \end{cases}\\
    0 &\mathrm{else}
    \end{cases}\,,
  \end{multline}
  characterized by the opening angle $\theta_{\mathrm{max}}$ of the ``cones'' that interact attractively between the particles,
  see also \Cref{fig:kf}.
Here, $\rv_{12}=\rv_{2}-\rv_{1}$ is the vector connecting the two centers of the particles, and $\hat{\rv}_i^{\alpha}$ is a unit vector from the  center  of  particle $i$ to  a  patch $\alpha$ on  its  surface,  depending  on  the  particle orientation $\Omega_i$. 
For a pair of patches one can define a bonding volume \cite{stopper2018bulk}
\be
 v_b = \frac{4\pi}{3}\, \sin^4\left( \frac{\theta_\text{max}}{2} \right)\,\left[ (\sigma +\delta)^3-\sigma^3\right] 
 \label{eq:bonding_vol}
\ee 
which is the angular-averaged volume integral over $\rv_{12}$ with unit contribution when the patches overlap.  
The patches are symmetrically placed at the corners of a tetrahedron and we choose a reference configuration for the location of the patches (for the unrotated state) by 
\begin{equation}
  P = \frac{1}{\sqrt{3}}\begin{pmatrix}
    -\sqrt{2} & \sqrt{2} & 0 & 0\\
    0 & 0 & -\sqrt{2} & \sqrt{2}\\
    1 & 1 & -1 & -1
  \end{pmatrix}\,.
\end{equation}
where each column in $P$ corresponds to one $\hat{\rv}_i^{\alpha}$.

In the following, the unit of length is the particle diameter $\sigma$ and the unit of energy is the square well depth $\epsilon$. We defined reduced particle number densities by $\rho^*=\rho\sigma^3$ and reduced temperatures by $T^*=\kT/\epsilon$. As usual, $\beta=1/(\kT)$. The parameters for the KF system are $\delta=0.119$ and $\cos\theta_\text{max}=0.895$, such that the single-bond-per-particle condition is always fulfilled. 

\subsection{Orientational expansions}
\label{sec:or_expansion}

The one-body density $\rho(x)$ in general depends on the spatial coordinates and the orientation of the molecule, $x=(\rv, \Omega)$. We write the density as a product of a position-dependent, orientationally averaged density and an orientational part, $\rho(\rv, \Omega)= \nav(\rv) \, \alpha(\rv,\Omega)$.  The orientational dependence is handled using an expansion into Wigner $D$-matrices,
\bea
  \alpha(\rv, \Omega) = \sum_{lmn} \alpha^{l}_{mn}(\rv) D^{l}_{mn}(\Omega) 
                         \equiv \sum_{(i)} \alpha^{(i)}(\rv) D^{(i)}(\Omega)\,, \label{eq:d_expansion}
\eea 
with orientational moments $\alpha^{(i)}$ where the set of indices $\{lmn\}$ is abbreviated by $(i)$. 

The tetrahedral symmetry of the KF patchy particle leads to constraints on the orientational moments. One can define an expansion which is explicitly invariant under tetrahedral group transformations:
\bea
 \alpha(\rv, \Omega) = \sum_{lmj} \alpha^{l}_{m[j]}(\rv) \Delta^{l}_{m[j]}(\Omega) 
                     \equiv  \sum_{(i')} \alpha^{(i')}(\rv) \Delta^{(i')}(\Omega)\,,
\eea
where the $\Delta^{(i')}$ are an extremely reduced set of linear combinations of $D^{(i)}$. Since the invariant functions $\Delta^{l}_{m[j]}$ are a linear combination of the Wigner D-functions $\sum_k d^{[j]}_k D^l_{mk}$, they are now labelled by the running index $[j]$ instead of a matrix index. The leading elements up to $l=6$ are given in Table \ref{tab:basis_fn}.

\begin{table*}

\begin{center} 
\begin{tabular}{@{}cc@{}cc@{}}
\toprule
  $l$ & basis& components & projection component \\  %
      & function &  & \\\midrule %
  3 & $\Delta^3_{m[1]}$ & $ \frac{1}{\sqrt 2} D^3_{m 2} + \frac{1}{\sqrt 2} D^3_{m \underline{2}}$  & $\frac{2}{\sqrt{2}}  D^3_{m 2}$ \\
  4 & $\Delta^4_{m[1]}$ & $ \frac{\sqrt 30}{12}D^4_{m \underline4} + \frac{\sqrt 30}{12} D^4_{m 4} -\frac{\sqrt 21}{6} D^4_{m 0}$ &$ - \frac{2\sqrt{21}}{7} D^4_{m0}$\\
  6 & $\Delta^6_{m[1]}$ & $ \frac{\sqrt 7}{4} D^6_{m \underline{4}} + \frac{\sqrt 7}{4} D^6_{m 4} + \frac{\sqrt 2}{4} D^6_{m 0} $ & $2\sqrt{2} D^6_{m0}$\\
      & $\Delta^6_{m[2]}$ & $- \frac{\sqrt{10}}{8} D^6_{m \underline{6}} + \frac{\sqrt{10}}{8} D^6_{m 6} -  \frac{\sqrt{22}}{8} D^6_{m\underline{2}} + \frac{\sqrt{22}}{8} D^6_{m 2} $ & $\frac{4 \sqrt{22}}{11} D^6_{m2}$\\
  \bottomrule
\end{tabular}
\end{center}
\caption{Linear combinations of Wigner matrices invariant under all tetrahedral group transformations up to $l=6$, note that $\underline{m}=-m$.
The last column shows the minimal Wigner $D$-matrix which is used  
to project out the corresponding moment from the simulation data.
(This follows from the symmetry of
the moments, e.g. $\alpha^3_{m2} = \alpha^3_{m \underline{2}}$.) 
}
\label{tab:basis_fn}
\end{table*}

Two-body functions like the interaction potential $v^\mathrm{KF}$, the pair correlation function $g$ and the direct correlation function $c$ depend on distance vector between the two points, $\rv=\rv_{12}$, and the orientations  $\Omega_1$ and $\Omega_2$ of the particle at the two points. They are expanded into orientational invariants
\bea
 f(\rv, \Omega_{1}, \Omega_{2}) = \sum_{m n l \mu\nu} f^{mnl}_{\mu \nu}(r) \Phi^{mnl}_{\mu \nu}(\Omega_{1}, \Omega_{2}, \hat{\rv}) 
 \label{eq:f_expansion}
\eea 
The basis functions $\Phi$ which are explicitly invariant under tetrahedral group transformations are given by
\begin{multline}
   \label{eq:Phi_Delta}
    \Phi^{mnl}_{[ji]}(\Omega_{1}, \Omega_{2}, \hat{\rv}) = c_{m}c_{n}\sum_{\mu' \nu' \lambda'} \tj{m}{n}{l}{\mu'}{\nu'}{\lambda'}  \\ \times 
   \conj{\Delta^{m}_{\mu' [j]}}(\Omega_{1}) \conj{\Delta^{n}_{\nu' [i]}}(\Omega_{2}) \conj{D^{l}_{\lambda' 0}}(\hat{\rv})
\end{multline}
using the standard (Wigner) 3$j$-symbols in round brackets. Further, $\overline{x}$ stands for the complex conjugate of $x$ and $c_n$ is a normalization constant with $c_n = \sqrt{2n + 1}$. Again, we use a short-hand notation 
$f^{(\alpha')} \equiv f^{mnl}_{[ji]}$. 
When evaluating the series expansion presented above, we need to chose a cutoff for the upper index $l$, or $m, n$ respectively, which we set to $l_\mathrm{max}=4$. 
Regarding the index combinations for the rotational invariants given by \eqref{eq:Phi_Delta}, the leading invariants up to $m=n=4$ are shown in Table \ref{tab:leading_gs}.
\begin{table}
\begin{center} 
\begin{tabular}{@{}lll@{}}
\toprule
  $m$ & $n$ & $l$ \\
\midrule
0 & 3 & 3 \\
0 & 4 & 4 \\
3 & 3 & 0, 2, 4, 6 \\
3 & 4 & 1, 3, 5, 7 \\
4 & 4 & 0, 2, 4, 6, 8 \\
\bottomrule
\end{tabular}
\end{center}
\caption{
Indices for the nonvanishing, independent components of two--body rotational invariants $\Phi^{mnl}_{[ij]}\equiv \Phi^{mnl} $ for $n_\mathrm{max} = 4$.
For the indices $[ij]$ there is only one possibility, thus they can be omitted. Consequently there are 16 independent components in total.
}
\label{tab:leading_gs}
\end{table}
\subsection{Simulations}

\begin{figure}
  \centering
  \includegraphics[width=.8\columnwidth]{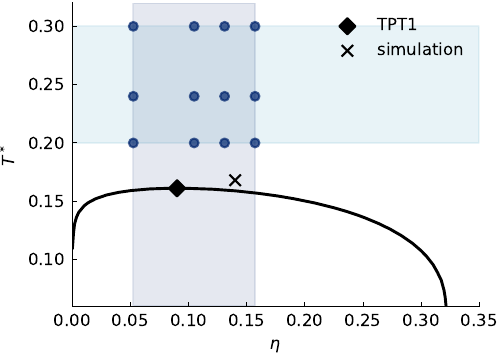}
  \caption{Phase diagram for the Wertheim theory based reference functional (TPT1, gas--liquid binodal:full line). The critical point from TPT1 is marked with a diamond, the one from simulations \cite{foffi2007possibility} by a cross. The blue dots indicate the state points under investigation.}
  \label{fig:phasediag}%
\end{figure}

{
We investigated pair correlations and density profiles between hard walls for a combination of three temperatures $T^*=\{0.20,0.24,0.30\}$ and four bulk densities $\rho^*=\{0.10,0.20,0.25,0.30\}$. 
} In Fig.~\ref{fig:phasediag}, the location of these points together with the gas--liquid binodal from TPT1 and the critical point from simulations \cite{foffi2007possibility} is shown in the $\eta$--$T^*$ plane where $\eta=\rho^*\pi/6$ is the packing fraction of the particles. 

Simulation data was gathered by Monte Carlo (MC) simulations 
in the NVT ensemble using the ``PatchyParticles'' code by Rovigatti et al.~\cite{rovigatti2018simulate}. In the code, a configuration is defined by  center positions and rotation matrices for all particles. The latter defines the orientation of particles with respect to an initial reference state.  
The number of particles was set to $N=1000$ and the length of
the simulation cube equal to $L = (N/\rho_\mathrm{avg})^{1/3}$, where $\rho_\mathrm{avg}$ is the average density in the box. 
The code was used in two variants.
For radial distribution functions $g$, we used an unmodified version of
the code with periodic boundary conditions. 
For obtaining the orientational moments $g^{(\alpha')}(r)$, averages over \num{10e7} snapshots were taken. For every snapshot, pairs of particles were sorted into distance bins with orientational index $(\alpha')$ and each pair configuration in a bin was weighted with the proper projection $D$-matrix (see last column of Table \ref{tab:basis_fn} and Eqs.~(\ref{eq:f_expansion},\ref{eq:Phi_Delta})), evaluated with the Euler angles for that configuration. This required the conversion of rotation matrices into Euler angles and fast evaluation of Wigner $D$-matrices. 

In the second version, the code was modified to exclude configurations of particles inside two parallel hard walls at variable distance (depending on the average number density). 
The averaged density profile and the orientational moments of the density distribution between the walls were obtained similar to the way described above for the pair correlations. 
While in the $NVT$ ensemble it is only possible to fix the average density $\rho_\mathrm{avg}$ in the simulation volume by changing $N$ or $V$,
it will be necessary to produce density profiles with certain fixed bulk values  $\rho_\mathrm{bulk}$  (i.e.\ far away from the walls) in order
to compare to DFT calculations.
We determined the correct average density by interpolating between a sufficient number of test points of the mapping
$\rho_\mathrm{avg} \leftrightarrow \rho_\mathrm{bulk}$.

\section{Integral equations}
\label{sec:ie}

\subsection{Theory}

Integral equation (IE) theory provides a scheme to obtain pair correlation functions $g$ and direct correlation functions (DCF) $c$ for a fluid with pair potential $\phi$. For the case of interest, $\phi=v_\text{HS}(|\rv_1-\rv_2|)+v^\text{KF}(x_1,x_2)$, the pair potential and $g(x_1,x_2)$ resp. $c(x_1,x_2)$ (with $x_i=(\rv_i, \Omega_i)$ denoting the combined position and orientation of particle $i$) depend on one distance ($|\rv_1-\rv_2|$) and five {Euler} angles in the bulk fluid.%
The integration over the $x_i$--variables is defined as $\int \odif{x_i} = \int \odif{\rv_i} \int \odif{\Omega_i}/K$, see Eq.~\eqref{eq:int_omega}. The Ornstein--Zernike (OZ) equation links $h=g-1$ and $c$ and is given by:
\begin{equation}
  h(x_1, x_2) = c(x_1, x_2) + \rho \int \odif{x_3} c(x_1, x_3) h(x_3, x_2)
\end{equation}
For solving the OZ equation, an additional closure relation is needed
\begin{multline}
    \ln \left[ h(x_1, x_2) + 1  \right] + \beta \phi(x_1, x_2) = \\ h(x_1, x_2) - c(x_1, x_2) + b(x_1, x_2)
    \label{eq:closure}
\end{multline}
with the unknown bridge function $b$. Having specified a model for the bridge function in terms of $h,c,\phi$, one can numerically solve the system of OZ and closure equation. 
A recent study on integral equations in this context, including comparison of different closure relations can be found in Ref.~\cite{pihlajamaa2024comparison}.
In the present work, the DCF is of special interest since it is needed as input in the molecular DFT route, see Sec.~\ref{sec:molecular_dft} below.

\paragraph*{HNC closure.}  The hypernetted chain (HNC) closure is defined by $b=0$. OZ and closure relation can be solved self--consistently without simulation input.

\paragraph*{MC/HNC closure.} The HNC closure is imposed beyond a cutoff $r_\mathrm{max}$, i.e. $\left.b(x_1,x_2)\right|_{r>r_\mathrm{max}}=0$. For $r \le r_\mathrm{max}$, the simulation result for the pair correlation function is imposed, i.e. $\left.g(x_1,x_2)\right|_{r \le r_\mathrm{max}}= g_\mathrm{MC}(x_1,x_2)$, requiring $r_\mathrm{max} < L/2$. This mixed closure allows to obtain the correct asymptotic behaviour of the  pair correlations at long distances and {thus} to get the correct  $q \rightarrow 0$-behaviour for both the spatial Fourier transforms $\hat g$ and $\hat c$. {Since HNC is asymptotically exact at long distances, this procedure gives  the exact $g$ and $c$ at all distances (within controlled numerical errors; see below).}

\paragraph*{Numerical method.} 
All correlation functions have been expanded into orientational invariants, see Sec.~\ref{sec:or_expansion}, specifically Eqs.~(\ref{eq:f_expansion},\ref{eq:Phi_Delta}) with $n_\mathrm{max} = m_\mathrm{max} = 4$. A principal problem is  the slow convergence of the expansion of $g$ which in the closure equation (\ref{eq:closure}) does not allow to reconstruct $g$ from the expansion {of} $\ln g$ and project out its expansion coefficients. Therefore, we employ the numerically elaborated method described in Ref.~\cite{belloni2017exact}. We also remark that the solutions of MC/HNC closure are not simply the HNC solutions for $r>r_\mathrm{max}$ since large and small distances mix in the OZ equation. A good criterion for a well chosen cutoff is the smooth behavior of $g$ and $b$ at the cutoff, i.e. specifically the bridge function should smoothly go to zero. In the actual calculations we set $r_\mathrm{max}=2.5$.  
The discontinuities in the KF-potential, both in radial and angular direction, require very small
mesh spacings for an accurate quadrature. In order
to reduce computing time we used a smoothed
KF-potential in the interval $r \in [1.099, 1.139 ]$,
keeping the second virial coefficient constant.
``Smoothing'' entails that the discontinuity is replaced by a linearly decreasing function in the above mentioned interval.

\subsection{Results}

\begin{figure}
  \centering
  \begin{subfigure}{.5\textwidth}
          \includegraphics[width=.9\columnwidth]{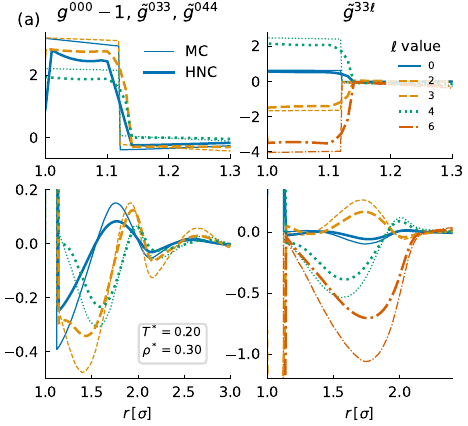}
  \end{subfigure} 
  \begin{subfigure}{.5\textwidth}
      \includegraphics[width=.9\columnwidth]{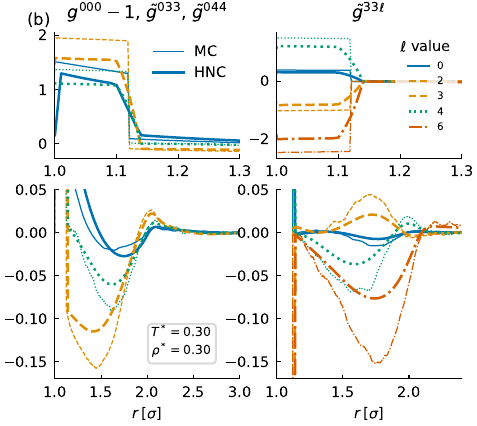}
  \end{subfigure}
  \caption{Moments of the pair correlation function $g$ for the state points $T^*=0.2$ (a), $T=0.3$ (b) and $\rho^*=0.3$. 
    Only the first few moments are shown here, the remaining ones can be found in the SI. All correlations except for $g^{000}(r)$ (blue curves) are scaled by the isotropic correlation $\tilde{g}^{\alpha'} = g^{\alpha'}/g^{000}$.
  Different linestyles/colors signify different $\ell$ values (see legend) and apply to all subgraphs. Thin lines are MC results, thick lines HNC results.}
  \label{fig:hnc_mc}
\end{figure}

We have determined  the "exact" MC/HNC  and approximate HNC pair correlations, direct correlation functions, and (in case of MC/HNC) the  bridge functions  in the supercritical region at the state points indicated in Fig.~\ref{fig:phasediag} ($T^*$ between 0.2 and 0.3 and $\rho^*$ between 0.1 and 0.3).
State points on the isotherm $T^*=0.2$ are already sufficiently close to the critical point ($T_c^* \approx 0.17$ and $\rho_c^* \approx 0.27$) that association leads to strong orientational correlations both around a test particle and near a hard wall \cite{simon2024machine} as well as to a noticeable density depletion close to the hard wall (drying).

In Fig.~\ref{fig:hnc_mc} we show results for $g$ from MC simulations (which enter MC/HNC) and compare to HNC for the state point $T^*=0.2$ and $\rho^*=0.3$ (lower right corner in Fig.~\ref{fig:phasediag}) where the association is strongest. 
Compared to MC, HNC underestimates the bonding (magnitude of $g^{000}$), the reduced magnitude of HNC solutions is also seen for all higher moments of $g$. However, this discrepancy almost disappears when the pure orientational contribution (defined by $g^{(\alpha')}/g^{000}$) is compared, as is done in Fig.~\ref{fig:hnc_mc}. The lower row  shows the same scaled moments  $g^{(\alpha')}/g^{000}$ for intermediate distances $r\in[1.0,3.0]$. Here, HNC qualitatively captures the oscillatory behavior of all moments but their magnitude is smaller compared to MC. 

For the highest temperature ($T^*=0.3$), strong orientational correlations are only seen in the bonding region. Similar to the lower temperature, HNC captures well the scaled moments $g^{(\alpha')}/g^{000}$. Beyond the bonding region, there is little structure in the first layer and beyond $r=2$ correlations are essentially zero. See also the SI for corresponding graphs.

The orientational moments of the bridge function (outside the core) have about the same magnitude as the moments of $g(r)$ and smoothly approach zero as $r \to r_\mathrm{max}$ which illustrates the self-consistency of the approach. {Again we refer to the SI for corresponding graphs.}

We postpone a detailed discussion of the solution for the DCF $c$ to Sec.~\ref{sec:molecular_dft}, where Molecular DFT is introduced which needs the DCF as its crucial ingredient.

\paragraph*{Equation of state.}

\begin{figure}
\centering
\begin{subfigure}{.5\textwidth}
  \centering
  \includegraphics[width=.85\linewidth]{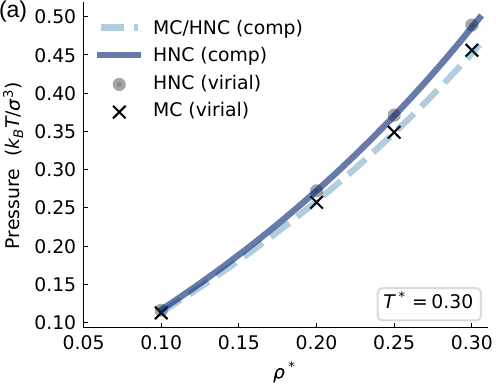}%
\end{subfigure}%
\newline
\begin{subfigure}{.5\textwidth}
  \centering
  \includegraphics[width=.85\linewidth]{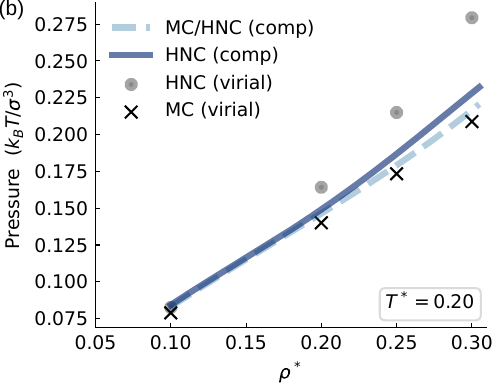}%
\end{subfigure}
\caption{Equation of state $p(\rho)$ for the two isotherms $T^*=0.2$ and 0.3, comparison between the compressibility and virial route for MC/HNC and HNC, respectively.
}
  \label{fig:ie3}%
\end{figure}

{We have computed the equation of state from pair and direct correlation functions using the standard compressibility and virial route. In the compressibility route}
\be
\label{eq:compressibility_route}
\beta p_\mathrm{com}(\rho) = \int_{0}^\rho \odif{\rho'}  \,(1 - \rho' \hat{c}^{000}(q=0;\rho')
\ee
The angular-averaged Fourier transform of the direct correlation function $\hat{c}^{000}(q;\rho)$ is provided by the IE approach, using either the HNC or the MC/HNC closure. The lowest available density point is $\rho_0^*=0.1$. To obtain the pressure for densities $\rho  \le  \rho_0$ we approximated $\hat{c}^{000}(q;\rho')$ in the integrand by $\hat{c}^{000}(q;\rho_0)$. Hence  
\begin{multline}
\label{eq:compressibility_route2}
\beta p_\mathrm{com}(\rho) = \\
\begin{dcases}
\rho \left(1 - \frac{1}{2} \rho \hat{c}^{000}(q=0;\rho_0) \right) & \rho \le \rho_0 \\
\beta p(\rho_0) + \int_{\rho_0}^\rho \odif{\rho'}  \,(1 - \rho' \hat{c}^{000}(q=0;\rho') & \rho > \rho_0
\end{dcases}
\end{multline}
From the virial route
\begin{multline}
\label{eq:p_virial}
\frac{\beta p_\mathrm{vir}}{\rho} = 1 + \frac{2\pi}{3} \rho \\
\times \left( g^{000}(\sigma_+) \sigma^3 + (g^{000}(\sigma'_+)-g^{000}(\sigma'_-)) {\sigma'}^3 \right)     
\end{multline}
where $\sigma'=\sigma+\delta$ is the range of the square well patch--patch potential in Eq.~(\ref{eq:sw_patch}).
Here one needs the value of the $g$ at the discontinuities; this introduces some systematic error due to the finite grid spacing. We expect this to be small for MC where we used a very fine grid but somewhat larger for HNC where the radial grid had a step size of \num{0.01} and the exact determination is difficult (as can be seen in \Cref{fig:hnc_mc}). In the actual computations, we used the virial pressure formula  for continuous potentials 
\begin{equation}
    \frac{\beta p_\mathrm{vir}}{\rho} = 1 - \frac{2 \pi}{3} \beta \rho \int g(r,\Omega) \pdv{v(r,\Omega)}{r} r^3 \odif{r}\odif{\Omega}
\end{equation}
which is consistent with the use of the smoothed KF-potential in the numerical solutions.

Results for $T^*=0.3$ and $T^*=0.2$ from the different routes (using MC and MC/HNC input) are shown in Fig.~\ref{fig:ie3}. 
The pressure isotherm for $T^*=0.3$ is dominated by hard sphere repulsions (at $\rho^*=0.3$, the
Carnahan-Starling pressure is $\approx 0.59$, thus attractions lower it by about 20 \%). 
The virial and compressibility route are consistent with each other within MC/HNC (as it should be) and HNC (not necessarily so), but HNC somewhat overestimates the pressure. 
For $T^*=0.2$ the effect of the patchy attractions is manifest in the downward deviation from the ideal gas behavior. 
At this lower temperature we see larger differences emerging between the two routes within HNC, with the values
determined by the compressibility equation are very close to those determined by simulation.
The virial route is influenced by both the internal inconsistency of HNC, and numerical errors due to the finite grid size or the smoothed potential.
\section{Density functional theory}
\label{sec:dft}

In DFT, the grand potential energy is a functional of the position- and orientation-dependent density $\rho(x)$:
\be
  \Omega[\rho] = \FE_{\mathrm{id}}[\rho] + \FE_{\mathrm{ex}}[\rho] + \int \odif{x} ( V^{\mathrm{ext}}(x) - \mu) \rho(x) \;,
\ee
and the equilibrium density $\req(x)$ is found by the minimum condition
\be
  \left. \frac{ \delta \Omega}{\delta\rho(x)} \right|_{\rho(x) =\req(x)} = 0 \;.
\ee 
The ideal gas functional $\FE_{\mathrm{id}}$ is given by
\be
  \label{eq:fid}
   \beta \FE_{\mathrm{id}} = \int \odif{\rv} \odif{\Omega}\, \rho(\rv, \Omega) \left[\ln(\rho \Lambda^{3} K) -1  \right] \,,
\ee  
with the thermal de--Broglie length $\Lambda$. The functional  $\FE_{\mathrm{ex}}$ is unknown in general, below we discuss two approaches for its construction.

\subsection{Molecular DFT}
\label{sec:molecular_dft}

Molecular DFT is based on a functional Taylor expansion of the free energy functional around a bulk (homogeneous) reference state (with density $\rho_0$). Using the density deviation $\Delta\rho(x)=\rho(x)-\rho_0$, this expansion can be written as:
\begin{multline}
  \label{eq:fex_mdft}
  \FE_\ex[\rho]  = F_\ex(\rho_0)  + \mu_\ex \int \odif{x} \Delta\rho(x) \\ - \frac{1}{2 \beta} \int \odif{x} \int \odif{x'} c(x,x';\rho_0) \Delta\rho(x) \Delta\rho(x') \\
+ \FE_B[\rho(x)] \;.
\end{multline}
 The first line is exact up to second order if the exact direct correlation function at the reference bulk state is used, here $F_\rmex(\rho_0)$ is the bulk free energy of the reference state {and $\mu_\rmex  = \frac{1}{V}\odv{F_\rmex}{\rho}(\rho_0)$ }. The functional $\FE_B$ (bridge functional) contains all higher orders in this expansion. For $\FE_B=0$, the full functional is termed HNC functional. Based on the approximate universality of the bridge function for simple fluids, one may approximate the bridge functional by the corresponding bridge functional for a hard sphere system \cite{oettel2005,Borgis2020}:
\begin{multline}
 \label{eq:fb_hs}
 \FE_B[\rho(x)] = \FE_{\ex,\hs}[\nav(\rv)] - F_{\ex,\hs}(\rho_0)  \\
   - \mu_{\ex,\hs}(\rho_0) \int \odif{\rv} \Delta\nav(\rv)  \\
   + \frac{1}{2\beta} \int \odif{\rv} \int \odif{\rv'} c_\hs(\rv - \rv';\eta_0) \Delta\nav(\rv) \Delta\nav(\rv') 
\end{multline}
where $\eta_0=\rho_0 \pi d^3/6$ is the packing fraction of hard spheres at the reference density and $d$ is a reference hard sphere diameter which can be used as a fit parameter. Furthermore, $\Delta\nav(\rv)=\nav(\rv)-\rho_0$ is the density difference of the averaged number density.
For actual calculations, {the fundamental measure theory (FMT), scalar  functional of Kierlik and Rosinberg was used~\cite{Kierlik1990,Kierlik1991}. In its Percus--Yevick version it was shown to be equivalent to the original Rosenfeld vectorial formulation~\cite{Rosenfeld1989,Phan1993}. It requires the use of the Percus--Yevick direct correlation function for $c_\hs$ in Eq.~(\ref{eq:fb_hs}). Note that a Carnahan--Starling version is also available~\cite{Kierlik1990}.}

In molecular DFT, the complete orientation dependence is contained in the second-order term through the direct correlation function $c$. It needs to be determined independently from IE for each reference state separately. The dependence on the reference state affects also the equation of state and the hard wall sum rule. We discuss this first for $\FE_B=0$, i.e.\ the HNC-functional. %
The HNC functional pressure $p(\rho)$ for a bulk density $\rho=\text{const}$ can be found from the compressibility rule 
\be
\label{eq:dpdbeta}
  \beta \odv{p}{\rho} = 1 - \rho \int \odif{\rv} c^{000}(r;\rho) \;  
\ee
%where the angular-averaged DCF is independent of $\rho$ and taken as the one at the reference density, $c^{000;\rho}(r)=c^{000}(r;\rho_0)$, and thus is independent of $\rho$. Integration gives
where for the angular-averaged DCF the one at the fixed reference density is taken, $c^{000}(r; \rho)=c^{000}(r;\rho_0)$. Thus integration of Eq.~\eqref{eq:dpdbeta} over $\rho$ gives
 $\beta p(\rho)  = \rho - \frac{\rho^2}{2} \int \odif{\rv} c^{000}(r;\rho_0) \;.$
Evaluated at the reference density, this defines a pressure which we call HNC-f pressure: 
\be
 \label{eq:phncf}
 p_\text{HNC-f}(\rho_0) =  \rho_0 - \frac{\rho_0^2}{2} \int \odif{\rv} c^{000}(r;\rho_0) \;. 
\ee 
If a hard sphere bridge functional is included, the angular--averaged DCF becomes
\be 
  c^{000}(r;\rho)=c^{000}(r;\rho_0) - c_\text{hs}(r;\rho) + c_\text{hs}(r;\rho_0) \;,  
\ee 
i.e. it acquires a density dependence through the hard sphere part. The pressure at the reference density becomes
\be 
 \beta p(\rho_0)=\beta p_\text{HNC-f}(\rho_0) - \beta p^\text{ex}_\text{hs}(\rho_0) + \frac{\rho_0^2}{2} \int \odif{\rv} c_\text{hs}(r;\rho_0)\;,
\ee 
where $p^\text{ex}_\text{hs}$ is the excess part of the hard sphere pressure. Inclusion of the bridge functional therefore changes the pressure at the reference density and thus also directly the contact density at the hard wall, as will be discussed in Sec.~\ref{sec:dft_results} below.

\begin{figure*}
\centering
\begin{subfigure}{\textwidth}
\centering
\includegraphics[width=0.8\columnwidth]{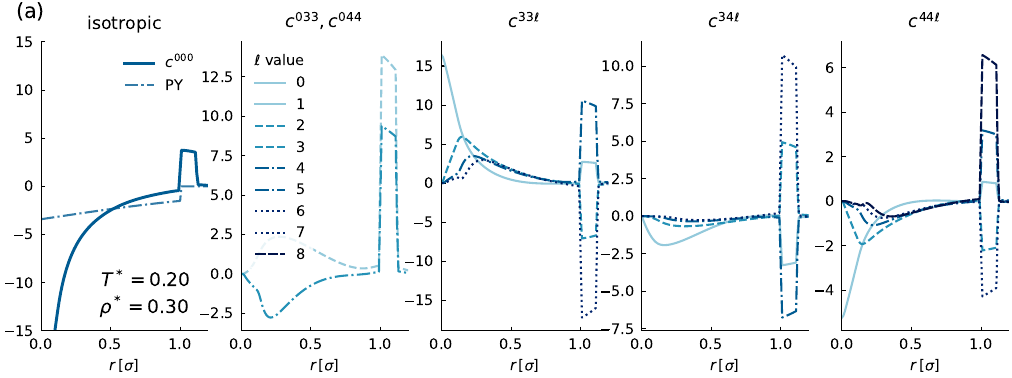}%
\end{subfigure}
\newline
\begin{subfigure}{\textwidth}
\centering
\includegraphics[width=0.8\columnwidth]{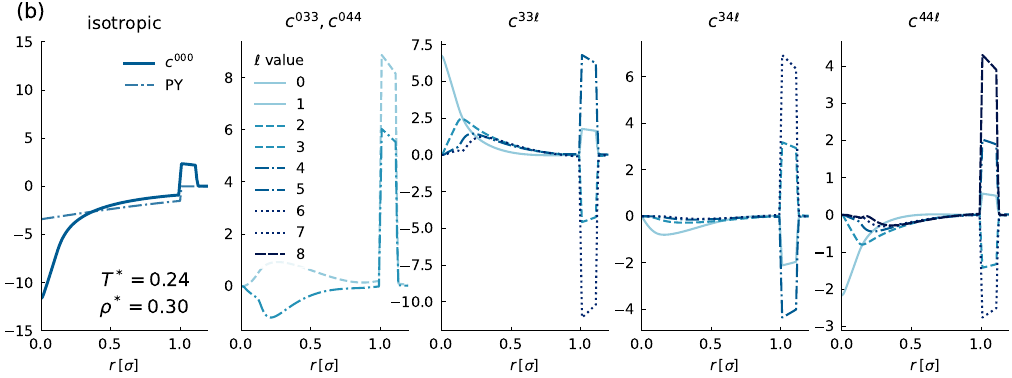}
\end{subfigure}
  \caption{
  Moments of the DCF $c(r)$ at density $\rho^*=0.3$ and temperatures $T^*=0.2$ (a) and 0.3 (b). The angular averaged (isotropic) moment $c^{000}$ is shown in comparison to the PY DCF. 
  }
  \label{fig:exact_c}
\end{figure*}

\begin{figure}
  \centering
  \includegraphics[width=0.9\columnwidth]{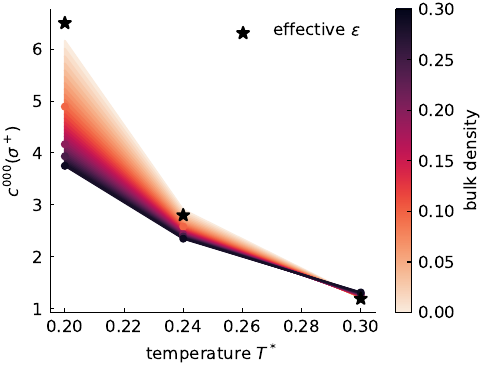}
  \caption{The shaded area shows $c^{000}(\sigma_{+})$ (shoulder peak in the bonding region) for different temperatures, 
  interpolated (cubic splines, for each temperature $T^*=0.2$, 0.24 and 0.3 individually) between the bulk densities $\rho \in [0, 0.3]$, where a lighter color corresponds
  to smaller densities. The upper edge shows the extrapolated low-density ($\rho \rightarrow 0$) limit which should correspond to Eq.~(\ref{eq:av_sw}). 
  }
  \label{fig:c000_shoulder}
\end{figure}
\paragraph*{Discussion of direct correlation functions.}
The crucial input for molecular DFT are the direct correlation functions determined in Sec.~\ref{sec:ie}.
The results for two state points (at density $\rho^*=0.3$ and temperatures $T^*=0.2$ and 0.3) are shown in Fig.~\ref{fig:exact_c}. The DCF moments are shown only in the distance range $r \in [0, 1.2]$ since they contain only contributions inside the hard core and in the bonding region $r \in [1, 1.119]$, and are essentially zero for larger distances. The isotropic component $c^{000}$ (left column) is shown together with the Percus--Yevick hard sphere DCF which is the high-temperature limit when the patchy attractions are unimportant. The shoulder-like peak of $c^{000}$ in the bonding region is due to the averaged patchy attractions and is discussed further below. 
Negative contributions in the core region $r < 1$ are usually attributed to repulsive interactions. Here we see that the patchy attractions at the lower temperature $T^*=0.2$ lead to a large negative enhancement as $r\to 0$ while for larger $r$ up to $r=1$ the negativity is decreased. At the higher temperature $T^*=0.3$ the deviation from the PY DCF are less pronounced, in line with the repulsion-dominated pressure at this temperature, see Fig.~\ref{fig:ie3}.
The higher moments $c^{(\alpha')}$ of the DCF have large contributions especially in the bonding region while the contributions in the core region are moderate (see the second to fifth column in Fig.~\ref{fig:exact_c}). 
The behavior in the bonding region shows a strong dependence on the 
index $\ell$ which modulates the dependence of the $c$-moment on the relative orientation between the particle at the center and the one at the distance $r$ (see definition of the basis functions in  Eq.~\eqref{eq:Phi_Delta}). Therefore it is also not surprising that 
the largest contributions to the core region ($r < 1$) result from the moments with $\ell = 0$ (cf.\ the columns
for $c^{33\ell}$ and $c^{44\ell}$) as there is no dependence on the relative orientation in this case.
This is different from the behaviour observed for the moments $c^{34\ell}$, as here the smallest possible value for $\ell$ is 1, %
hence a dependence on the relative orientation is always present. 

The shoulder-like peak in the bonding region of $c^{000}$ shows a strong dependence on the density as $\rho \to 0$. In this limit, the strength is related to an angular-averaged interaction between the particles which is an isotropic square well potential of the same range and strength $\epsilon'$. Imposing the same contribution from attractions to the second virial coefficient in the patchy and the averaged model, we find 
\cite{foffi2007possibility}
\begin{gather}
    c^{000}(\sigma_{+}) \approx \exp(\beta \epsilon') -1  = \chi^2 (\exp(\beta \epsilon) -1 ) \\
    \chi =(1-\cos\theta_\text{max})\np/2 \;. 
  \label{eq:av_sw} 
\end{gather}
Extrapolating our data for the strength of the shoulder to $\rho \to 0$ at the three temperatures 0.2, 0.24 and 0.3 agrees quite well with this limit (see Fig.~\ref{fig:c000_shoulder}) and illustrates the strong density dependence, especially for $T^*=0.2$. 
\subsection{Machine learning approach}

\subsubsection{Splitting of the density functional}

We follow the approach introduced in Ref.~\cite{simon2024machine} and split the excess free energy into a reference part, depending only on the averaged density profile $\nav(\rv)$, and an orientational part,
\begin{equation}
  \label{eq:fexsplitting}
\FE_{\text{ex}} = \FE_{\text{ex,iso}}[\nav(\rv)] +  \FE_{\text{ex,or}}[\nav(\rv), \alpha(\rv, \Omega)] \,,
\end{equation}
For the reference part, the ``Stopper--Wu'' functional was used \cite{stopper2018bulk} whose equation of state equals the Wertheim TPT1 result for associating fluids, for the associated gas--liquid binodal see Fig.~\ref{fig:phasediag}. The TPT1 equation of state gives a fair approximation in the whole phase diagram but differences especially in the supercritical region are visible, which are manifest in the contact value of density profiles at hard walls. 
Regarding $\FE_\mathrm{ex,or}$, in Ref.~\cite{simon2024machine} a generalized mean-field density functional 
was constructed via ML methods which describes the orientational correlations fairly well for higher temperatures ($T \lesssim 1.5 T_c$) but shows deficiencies for temperatures closer to $T_c$ ($T \sim 1.2 T_c$). The ML-functional has a mean-field kernel which is much stronger than expected from a naive RPA approach (expansion of the orientation dependent part of the KF potential in orientational invariants). The ML functional has been constructed for inhomogeneous profiles in the flat wall geometry, and therefore it cannot be applied straightforwardly to other inhomogeneous situations.

\paragraph*{Tuning of the FMT kernels in $\FE_{\mathrm{ex,iso}}[\nav]$.}
In the following, we use ML with hard wall simulation data as training input to improve the Stopper--Wu reference functional. 
Specifically, we tune the convolution kernels (weight functions) used in the \emph{bonding} part.
Conceptually, this type of training of an analytic form for the density functional has been pursued previously for systems in 1D in Refs.~\cite{shang2019classical,shang2020FEQL} and in 3D in Ref.~\cite{cats2021}.
The Stopper--Wu excess free energy consists
of the hard-sphere and bonding contribution \cite{stopper2018bulk,simon2024machine} 
\begin{equation}
  \label{eq:fref}
\beta \FE_\text{ex,iso}[\nav] = \int \odif{\rv}  \left[ \phi_\hs(\{ n_\nu\}) + \phi_\bond( \{ n_\nu\} ) \right]
\end{equation}
both depending on the same set of weighted densities $n_\nu$. Since the original Rosenfeld functional for the hard-sphere part is already very accurate we do not want
to modify the weights that enter $\phi_\hs$, but only those that appear in 
$\phi_\bond$. The latter is given by
\begin{multline}
\phi_{\mathrm{bond}}(\left\{ n_{\nu}(\rv) \right\}) = \np n_{0}(\rv) \xi(\rv)^{3} \\ \times
\left[ \ln X(\rv) - \frac{X(\rv)}{2} + 1/2 \right] \,.
\end{multline}
Here, the weighted density $n_0=\nav \otimes w_0$ is a convolution ($\otimes$) of the density with the weight function $w_0(\rv)=\delta(R-r)/(4\pi R^2)$. Further, $\xi(\rv)=1-\ntwo \cdot \ntwo/n_2^2$ with the weighted densities $\ntwo$ resp. $n_2$ utilizing the weight functions $\wtwo=\delta(R-r) {\mathbf e}_r$ and $w_2=\delta(R-r)$. Finally, $X(\rv)$ is a position--dependent bonding probability, defined through the solution of the equation
\be
  \frac{X(\rv)}{(1-X(\rv))^2} = \np\,n_0(\rv)\,\xi(\rv)^3\, \Delta( n_2,\ntwo,n_3)\;.
\ee
Here, $\Delta= v_b(\exp(\beta\epsilon)-1)g_\text{hs}( n_2,\ntwo,n_3)$ depends on the bonding volume $v_b$(Eq.~(\ref{eq:bonding_vol})), the patch attraction strength $\epsilon$ and a  generalized contact value of hard spheres, given by
\be
   g_\text{hs} = \frac{1}{1-n_3(\rv)} + \frac{\sigma\,n_2(\rv)\, \xi(\rv)^3}{4(1-n_3(\rv))^2} + 
      \frac{\sigma^2\,n_2(\rv)^2\,\xi(\rv)^3}{72(1-n_3(\rv))^3} \;.
\ee 
where the last weighted density
$n_3$, with weight function $w_3(\rv)=\theta(R-r)$ enters.
This completes the definition of $\phi_\text{bond}(\{n_\nu\})$ which is seen to depend on the set of weighted densities $\{n_\nu\}=\{n_0,n_2,n_3,\ntwo\}$ with the corresponding set of weight functions $\{w_\nu\}=\{w_0,w_2,w_3,\wtwo\}$. In our ML approach, it is this set of weight functions which will be learned from the hard wall simulation data.
We denote the improved weighted densities by $m_\nu$
\begin{equation}
    m_\nu = \rho \otimes w'_\nu = \rho \otimes \left( w_\nu  +  \varpi_\nu \right)
\end{equation}
They result from a convolution of the density with the usual FMT weight $w$, corrected
by an additional term $\varpi$, which is determined during training. If we denote the 
free parameters in the parametrization of $\varpi$ by $\xi$, the complete excess free energy reads
\begin{equation}
\FE_\ex[\rho] = \FE_\hs[\{ n_\nu \} ] + \FE_\bond[ \{ m_\nu(\xi) \}]
\end{equation}

\paragraph*{Full functional in flat wall geometry.}
In the flat wall geometry, the density profile only depends on the distance $z$ from the wall, besides the orientation, i.e.\ $\nav \equiv \nav(z)$ and $\alpha^{(i')} \equiv \alpha^{(i')}(z)$. In the bonding part of the reference functional, the ML-weighted densities likewise depend only on $z$ and can be written as a convolution with regard to $z$ only, $m_\nu(z) = \int dz' \omega_\nu(z') \rho(z-z')$. The relation of $\omega_\nu$ to $w'_\nu$ for the scalar weights $(\nu=0,2,3)$ is given by
\be
 \label{eq:z_to_r}
  \omega_\nu(z) = 2 \pi \int^\infty_z \odif{r}\,  r\, w'_\nu(r) \;.%
\ee 
For the vector weight, we set $\wtwo'(\rv) =  {\mathbf e}_r \tilde w_2'(r)$ which results in 
$\ntwo(z)={\mathbf e}_z \int dz' \tilde \omega_2 (z') \rho(z-z')$ with
\be
 \label{eq:z_to_r2}
  \tilde \omega_2(z) = 2 \pi z \int^\infty_z \odif{r}\,  \tilde w_2'(r) \;.
\ee
In ML, actually the difference of the weights $\omega_0,\omega_2,\omega_3,\tilde \omega_2$ to the corresponding Rosenfeld weights are learned from which the general weight functions can be reconstructed.

The orientational part of the functional in flat wall geometry is the ML mean field functional of Ref.~\cite{simon2024machine}:
\begin{multline}
  \label{eq:mf_ml}
    \FE_{\text{ex,or}}^\text{mf} = \frac{A}{2} \int \odif{z}\odif{z'} \nav(z) \nav(z') \\ \times
    \sum_{\substack{i,j \\ \text{not }i=j=0}} \alpha^{(i)}(z)   \alpha^{(j)}(z') M^{ij}(z- z') \,,
\end{multline}

where $A$ is the wall area. The truncated expansion of $\alpha(z,\Omega)$  is given by
\be
 \alpha(z,\Omega) = \alpha^{(0)}(z) + \alpha^{(3)}(z) \,\Delta^3_{0[1]}(\Omega) + \alpha^{(4)}(z) \,\Delta^4_{0[1]}(\Omega) \;,
\ee 
see Tab.~\ref{tab:basis_fn} and Ref.~\cite{simon2024machine}.
The moments $M^{ij}(z)$ are the mean--field kernel moments determined by ML.  The absence of $M^{00}(z)$ in the sum ensures that there is no contribution of $\FE_{\text{ex,or}}^\text{mf} $ in the homogeneous bulk, as there the free energy is determined from the reference functional only. In contrast to the reference part, here the knowledge of $M^{ij}(z)$ is not sufficient to reconstruct the mean field kernel applicable to all geometries, i.e. to inhomogeneities other than flat wall geometries, see also the discussion in Ref.~\cite{simon2024machine}.  

\subsubsection{Training the weight functions for the reference part}

The kernel parameters are trained iteratively, by solving the Euler--Lagrange equation for the density profile
$\hat{\rho}_\xi$ at the present parameters $\xi$ and then computing the gradient using backpropagation with respect to the loss
\begin{equation}
    L(\xi) = \sum_i \left( \reqs(z_i) - \hat{\rho}_\xi(z_i) \right)^2
\end{equation}
where $\reqs(z_i)$ are the equilibrium values determined by simulations.
This leads to self-consistent weights throughout the procedure.
We observe however, that the procedure tends to break down after a certain period of time, probably
because the previous gradient step produced parameters that do not allow for a self-consistent fixed point.
We deal with this by using the last set of parameters that produced consistent results.

Similarly we could observe that the range of the support of the augmented kernels influences the
stability of the training procedure. 
Smaller values of the kernel range tended to be less stable numerically than large ones. On
the other hand the range should not be chosen too large since we know that 
the Kern--Frenkel interaction is limited to a radius between the values 1 
and 1.119.
We empirically determined a good trade-off to be at $2$.
Training was done using \num{35} density profiles between two hard walls at variable distance (at state points specified by $\rho^*_\text{avg} \in [0.1, 0.13, 0.16, 0.19, 0.21, 0.24, 0.27]$ and $T^* \in [0.2, 0.22, 0.24, 0.26, 0.30]$), with a resolution $\Delta z=L/1024$.

\begin{figure}
  \centering
\begin{subfigure}{.5\textwidth}
  \centering
\includegraphics[width=.9\linewidth]{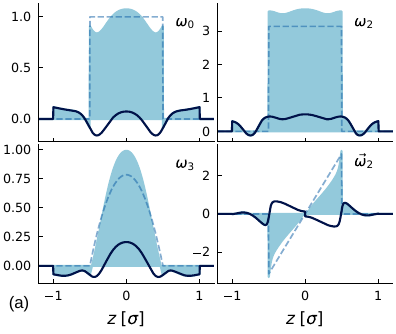}%
\end{subfigure}%
\newline
\begin{subfigure}{.5\textwidth}
  \centering
  \includegraphics[width=.9\linewidth]{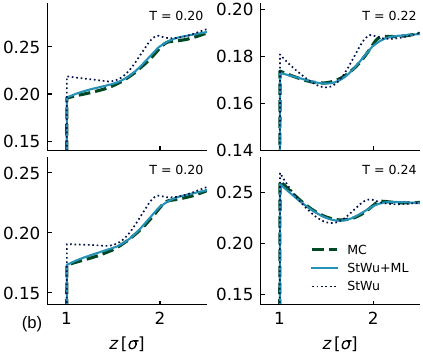}%
\end{subfigure}

  \caption{Improved kernels and the resulting density profiles. (a) The original FMT kernels are shown in as dashed lines, the learned $\varpi_\nu$ as thick lines and the resulting improved kernels as colored area.
  (b) Density profiles from simulations (MC), the Stopper--Wu functional (StWu) and the ML--augmented Stopper--Wu functional (StWu+ML) at the following ($T^*,\rho_\text{bulk}^*$) state points: (0.2, 0.28), (0.2, 0.25), (0.24, 0.24), (0.22, 0.19) from upper left counterclockwise.}
  \label{fig:impkernel}
\end{figure}

The resulting kernels are shown in Fig.~\ref{fig:impkernel}, as well as associated density profiles (from the training set). The description of the training data is very satisfactory, note that the augmented weight functions $m_\nu(z)$ are temperature--independent and thus cover the range from weak to strong bonding when $T^*$ is varied between 0.3 and 0.2.

\subsubsection{Direct correlation functions}

\begin{figure}
\centering
  \includegraphics[width=\linewidth]{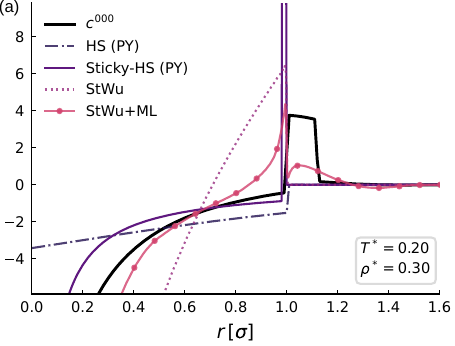}%
\newline
  \includegraphics[width=\linewidth]{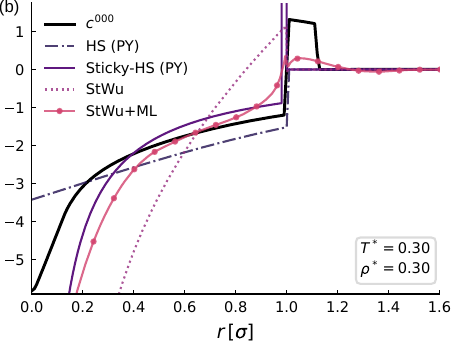}%
\caption{Comparison between direct correlation functions from MC simulations ($c^{000}$, angular averaged), hard sphere Percus--Yevick (PY), sticky hard spheres (PY), the Stopper--Wu functional (StWu) and the ML--augmented Stopper--Wu functional (StWu+ML) at density $\rho_\text{bulk}^*=0.3$ and two temperatures $T^*=0.2$ and 0.3. 
}
\label{fig:dcf}
\end{figure}

Given the improved weighted density kernels we can compute 
the corresponding direct correlation function $c_\text{iso}(r)$, which 
becomes
\begin{equation}
    \label{eq:c2_conv}
    c_\text{iso}(r) = - \sum_{ij} \pdv[delims-eval=.|]{\phi_\bond}{m_i, m_j}_\rhob \left( \pm w'_i \otimes w'_j \right) (r)  + c^\hs(r)\,,
\end{equation}
where the minus sign holds when vector-like weighted densities are involved.
The result is shown in Fig.~\ref{fig:dcf}, together with the original Stopper-Wu DCF, the angular averaged DCF $c^{000}$ from simulations, the hard sphere PY DCF and the sticky hard sphere PY DCF.
This allows for the following interesting observations.
The DCF of the Stopper--Wu reference functional shows a rather strong (integrable) divergence for $r \to 0$ coming from the bonding part, for the mathematical reasons see Ref.~\cite{lutsko2013_pre_87_014103}. {A similar divergence is found in the PY solution for the sticky hard sphere model \cite{baxter1968} which is also shown in Fig.~\ref{fig:dcf}. For the sticky hard spheres the stickiness parameter is chosen such that the contribution from attractions to the second virial coefficent is equal to the one in the patchy model.} The MC/HNC data for $c^{000}(r)$ do not show this divergence but a negative enhancement, as discussed in the context of Fig.~\ref{fig:ie3}. The DCF from the ML--augmented functional still shows the divergence as $r \to 0$ but for finite $r$ the curve is much closer to the angular averaged DCF from simulation. {Overall, the behavior inside the core of $c^{000}(r)$ and the DCF's from StWu+ML and the Percus--Yevick solution for sticky hard spheres is very similar.}    
A second effect of the patchy attractions in $c^{000}(r)$ is the reduction of the negative part as $r \to 1$ and the positive shoulder in the bonding region.
In the DCF of the Stopper--Wu reference functional, this effect is crudely mimicked by a ramp-like part due to attractions (as $r\to 1$) which however is  confined to  $r < 1$ which is a consequence of the limited range of the FMT weights. The ML--augmented functional has a DCF with a reduced strength of this ramp-like part and thanks to the larger range of the kernels shows a bonding contribution outside of the hard core. 
The structure of the improved DCF is restricted by the functional derivatives of the excess free energy in~\eqref{eq:c2_conv}, which remains untouched.
Nevertheless, the improved kernels produced a much better approximation of the exact, angular-averaged DCF, without
having been trained on this objective.

\subsection{Results}
\label{sec:dft_results}

In the following, we will compare results from different density functionals:
\begin{itemize}
  \item MDFT-HNC: the free energy (\ref{eq:fex_mdft}) of molecular DFT with no bridge functional ($\FE_B=0$) but with the full direct correlation function $c$ from Sec.~\ref{sec:ie} (MC/HNC). In IE language , this corresponds to a binary system of KF (species 1) and ``wall'' particles (species 2) in the dilute limit of species 2 and an HNC approximation for the correlations between species 1 and 2.
  \item MDFT-HSB:  the free energy (\ref{eq:fex_mdft}) with the hard sphere bridge functional (\ref{eq:fb_hs}) and the full $c$ from MC/HNC.
  \item MDFT-HSBav/MDFT-HNCav: the free energy (\ref{eq:fex_mdft}) with and without, respectively, the hard sphere bridge functional (\ref{eq:fb_hs}) and the $c$-function reduced to the orientationally averaged $c_\text{av}=c^{000}_{[00]}$ from MC/HNC.
  \item StWu+ML: the ML improved Stopper--Wu reference functional (\ref{eq:fref})
  \item StWu+MF: {the Stopper--Wu reference functional and the mean--field ML functional (\ref{eq:mf_ml}) with kernels as determined in Ref.~\cite{simon2024machine}}
\end{itemize}

\subsubsection{Contact density at hard wall and equation of state}

\begin{figure}
\centering
\begin{subfigure}{.4\textwidth}
  \centering
  \includegraphics[width=\linewidth]{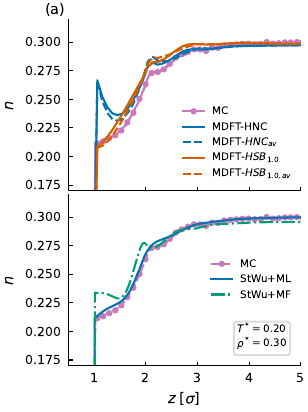}
\end{subfigure}%
\newline
\begin{subfigure}{.4\textwidth}
  \centering
  \includegraphics[width=\linewidth]{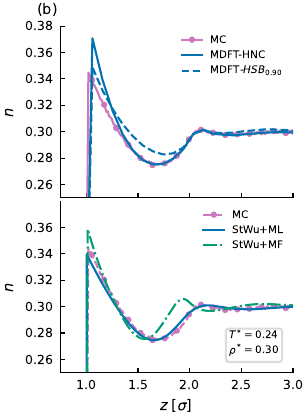}
\end{subfigure}
\caption{Averaged density profiles from the different functionals at two state points $\rho^*=0.3$ and
$T^*=0.2$ (a) and $T^*=0.24$ (b). The upper row shows results from MDFT, and the lower row from the reference functional route. For MDFT-HSB, 
the subscript indicates the reference HS-fluid diameter, that was adjusted to reproduce the correct KF-fluid bulk pressure at the temperature $T^*$.}
\label{fig:density_wall}
\end{figure}

{According to the contact theorem, the bulk pressure of the fluid at bulk density $\rho_0$ is given by
\be
 \label{eq:p_contact}
 \beta p_\mathrm{hw} (\rho_0) = \nav(z_\text{w}^+)
\ee
where $\nav(z_\text{w}^+)$ is the contact density of the fluid at a hard wall at location $z_\text{w}$. 
The contact density and thus the pressure can be simply read off the MC simulation profiles and it is consistent with the pressure from IE using the MC/HNC closure.
}
In Fig.~\ref{fig:density_wall}, the average density profiles at the hard wall are shown for the different approaches for a bulk density $\rho^*=0.3$ and temperatures $T^*=0.2$ and 0.24. Upper panels show MDFT results, lower panels results from the reference functional route. Regarding the contact density $\rho_c=\nav(z_\text{w}^+)$ one sees that only MDFT-HNC gives deviating results from the MC results which are systematically too large.  
{The MDFT-HNC contact value is related to the pressure $p_\text{HNC-f}$ defined in Eq.~(\ref{eq:phncf}) which is deficient since it is based on the approximation $c^{000}(r;\rho<\rho_0)\approx  c^{000}(r;\rho_0)$ in the integration of the compressibility rule.} The inclusion of a hard sphere bridge functional can correct this, for $\rho^*=0.3$ and $T^*=0.2$ we have chosen $d=1$ as the reference diameter which brings the contact density to the simulation value. For $\rho^*=0.3$ and $T^*=0.24$, the inclusion of the bridge functional also brings the contact value closer to the simulation value but at the same time leads to an increased overall deviation of the profile. The contact densities for the averaged and orientation-resolved functionals of the same type (i.e. MDFT-HNC vs. MDFT-HNCav, MDFT-HSB vs. MDFT-HSBav and StWu+ML vs. StWu+MF) turn out to be the same.

\subsubsection{Averaged density \texorpdfstring{$\nav(z)$}{n} near the hard wall}

Regarding the full profile $\nav(z)$, the ML-fitted reference functional gives the best results and clearly improves the results from the original Stopper-Wu reference functional (StWu in fig.~\ref{fig:density_wall}). 
In MDFT, the inclusion of the hard sphere bridge functional leads to a satisfactory description of the onset of drying (see the results for $\rho^*=0.3$ and $T^*=0.2$) and there is only a small difference between the averaged and orientation-resolved functionals.

In a recent study on associating, patchy fluids~\cite{barthes2024molecular}, a new density functional for the averaged density was introduced
which is similarly based on Wertheim’s thermodynamic perturbation theory, utilizing an inhomogeneous 
formulation with certain approximations for inhomogeneous pair distribution functions.  With this new
functional, accuracies similar to those of $i$SAFT~\cite{tripathi2005microstructure} are obtained. 
While the density profiles at a hard wall resulting from this approach seem to be slightly worse than
those obtained in the present work, the functional can more easily applied to different geometries and even complex
three-dimensional problems but at less computational cost.

\subsubsection{Orientational moment profiles \texorpdfstring{$\alpha^{(i)}(z)$}{α(z)} near the hard wall }

\begin{figure} 
\centering
  \includegraphics[width=.9\linewidth]{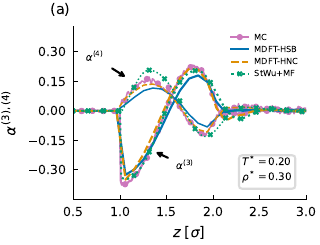}
  \includegraphics[width=.9\linewidth]{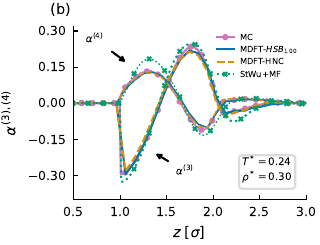}
\caption{Orientational moments from the different functionals at two state points $\rho^*=0.3$ and
$T^*=0.2$ (a) and $T^*=0.24$ (b). The upper row shows results from MDFT, and the lower row from the reference functional route.
}
   \label{fig:alpha_wall}
 \end{figure}

Fig.~\ref{fig:alpha_wall} shows the results for the leading two orientational moments of the orientation--resolved density profile at the hard wall for the same two state points as before. Here MDFT-HNC gives a very accurate description of the simulation data. The inclusion of the bridge functional leads to only small changes in the $\alpha^{(i)}(z)$, despite the fact that the averaged density profile is altered considerably. Both MDFT-HNC and MDFT-HSB give a very good description also for the other state points investigated (not shown). Thus, the bulk--density dependent DCF's are sufficient to capture the orientational correlations in this system. 

For the reference functional StWu+MF, the description of the $\alpha^{(i)}(z)$ is qualitatively correct but with some quantitative differences, {especially when compared with MDFT-HNC}.
Both StWu+MF and MDFT-HNC are mean--field like (quadratic in the orientation--resolved density profile) such that the explicit density dependence of the DCF (the kernel in MDFT-HNC) suggests that the bulk--density independent \textit{ansatz} for the ML mean-field kernel from Ref.~\cite{simon2024machine} is not {fully} sufficient.

 The results of \Cref{fig:density_wall,fig:alpha_wall} imply that the effect of lowering the temperature at a fixed density is first to orient the particles at the wall (with one patch pointing perpendicular to the wall, see also the visualizations in Ref.~\cite{simon2024machine}) and upon further decrease of temperature, the region near the wall becomes depleted (onset of drying).

\subsubsection{Hard spherical tracers}

\begin{figure*}
  \centering
  \includegraphics[width=.8\linewidth]{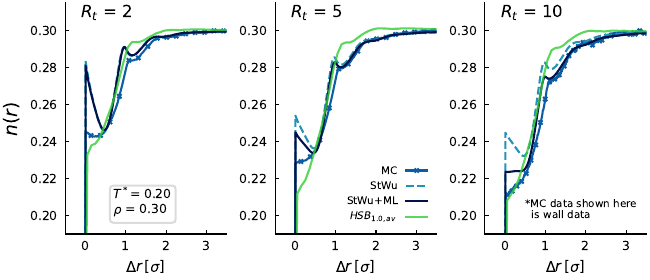}
  \caption{Orientation averaged density profiles $n(r)$ at bulk density $\rho^* = 0.30$ and temperature $T^*=0.20$ around a hard tracer particle with varying exclusion radius $R_t$ at distance $\Delta r$ from the tracer. In the limit $R_t \rightarrow \infty$ the hard-wall geometry is recovered.
}
  \label{fig:tracer}
\end{figure*}
\begin{figure}
  \centering
\includegraphics[width=.8\columnwidth]{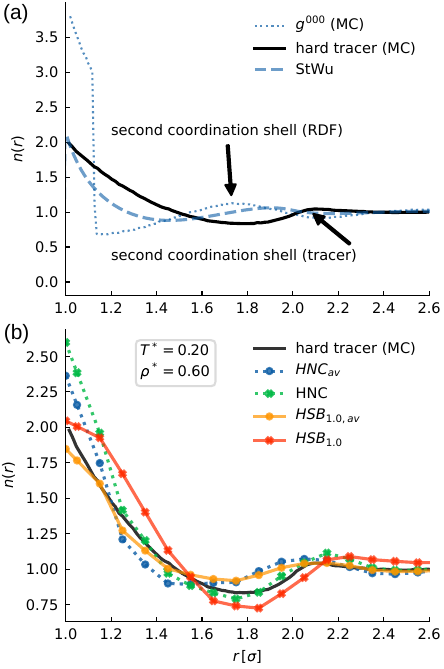}
\caption{(a) Normalized density profile of a patchy particle liquid around another patchy particle (i.e.\ the RDF) and around a hard spherical tracer of radius $1\sigma$, with corresponding shift of the position of the second coordination shell. Further the DFT density profile resulting form the Stopper--Wu functional is shown. (b) Normalized density profiles of patchy particles around a hard tracer with radius $1\sigma$ resulting from different MDFT functionals.}
  % \caption{Bulk density profiles of patchy particles around (i) another patchy particle (patchy RDF) and (ii) around a hard tracer (hard sphere) and the corresponding DFT results from the StWu and the MDFT functional.}
  \label{fig:peak_compare}
\end{figure}

It was observed in previous studies that the predictions
of the Stopper--Wu functional around spherical tracers miss
the correct positions of the second coordination shell~\cite{stopper2018bulk}.
The reason for this is the  suppression of tetrahedral bonds in the first shell around the tracer,
due to its spherically symmetric interaction. As a consequence the second peaks shifts further away
from the tracer particles' center to a distance of about $1\sigma$ (see \Cref{fig:peak_compare}, this example for a rather high bulk density $\rho^*=0.6$ is further discussed below).
In the following we investigate whether the ML-improved and the MDFT functional are able to
predict this structural property better.
In order to assess the performance around the tracer, we examine the structure around 
a radial tracer with varying radius $r$, where for large values of $r$ we reproduce the situation
of the hard-wall. In this way we can gradually go from one regime to the other. Finally, 
we test the high density case, where the shifted-peak phenomenon manifests itself the strongest.

We have examined the orientationally averaged density profile $n(r)$ at an average density $\rho^* = 0.3$ and temperature $T^*=0.20$ around hard spherical tracers with exclusion sphere radius values 2, 5 and 10 see Fig.~\ref{fig:tracer}. 
The ML-enhanced Stopper--Wu functional can also be minimized in this geometry since the weights $w_\nu(z)$ in wall geometry can be converted into weights  $w'_\nu(r)$ in radial geometry using Eqs.~(\ref{eq:z_to_r}, \ref{eq:z_to_r2}). 
It is seen that ML has essentially ``learned'' the drying effect which for the largest tracer is similar to the hard wall, whereas the effect seems to disappear for smaller tracers.
For the smallest tracer, results from the original and the ML-enhanced Stopper-Wu functional are essentially indistinguishable.
On the other hand, MDFT shows
an improved value of the contact density using the HSB functional, even without taking 
orientations into account, i.e.\ using only the averaged number density.
The HNC functional alone produces profiles that are similar to the StWu results, and are therefore not
shown in \Cref{fig:tracer}.
While the contact value is relatively well described by the HSB functional, there are still some deviations (of few percent) w.r.t.\ the fluid structure  
close to the wall (the region $r \in [1, 3]$),
presumably because correlation between the patchy particles is underestimated leading to drying film that is too thin.
The same behavior is also observed for the density profiles at the wall with strong drying, see \Cref{fig:density_wall}.

For the high bulk density case (at $\rho^* = 0.6$) corresponding results are show in \Cref{fig:peak_compare}).
The second peak in the density profile (the second coordination shell) is shifted to a distance from the tracer somewhat larger than 1 and appears to be hard-sphere like.
In contrast, the second coordination shell seen in the averaged pair correlation function $g^{000}(r)$ is at a smaller distance of about 0.75 (``tetrahedral network peak''), i.e.\ hard tracer breaks the tetrahedral network.
The shift of the second coordination shell peak is captured better by MDFT than by the Stopper--Wu potential, as can be seen in \Cref{fig:peak_compare}, though differences remain.
The contact density is well captured by the anisotropic HSB functional, though here the structure is slightly off. The HNC closure (with fully resolved anisotropy) produces
a better structural prediction, but
contact values that are off by about 25\%.

In contrast to the cases at the hard wall, orientational information
becomes more important for accurate predictions in the MDFT approach.

\section{Summary and conclusions}
\label{sec:summary}
In this work, we have discussed the orientational structure of a 4-patch Kern--Frenkel bulk fluid and developed orientation-resolved density functionals for the general inhomogeneous problem. For the bulk structure, we have computed fully angle-resolved pair and direct correlation functions using a combination of Monte Carlo simulations and integral equation approaches. For several supercritical state points the orientational expansion of the pair correlations in symmetry-adapted base functions reveals strong orientational correlations which, however, decay quickly beyond a range of two hard diameters. The direct correlation functions are a necessary input for the molecular DFT approach to construct a density functional. They show an explicit bulk density and temperature dependence which is important to capture the orientational correlations near hard walls. As a second approach to a functional, we considered the reference functional approach where the reference functional only depends on the angular-averaged density and the remainder is parameterized in mean-field form. The reference functional is trained by machine  learning methods, based on an analytic form of ``functionalized'' Wertheim theory. This analytic form allows a training of the functional in wall geometry which then can applied in other geometries as demonstrated for hard spherical tracers. 

The functionals have been minimized in the presence of hard wall inhomogeneities which is a stringent test since near the hard wall both the onset of drying and strong orientational ordering is observed. Angular-averaged functionals based on molecular DFT and the reference functional show similar performance with the reference functional being more accurate over the range of investigated state points. With regard to the description of orientational correlations, molecular DFT gives an excellent description but relies on the computationally intensive determination of the direct correlation functions. 

Based on our work, we can formulate some perspectives for the use of machine learning techniques for finding density functionals for anisotropic particles which are applicable to general inhomogeneities. This problem clearly goes beyond the present state-of-the art of ML enhanced classical DFT, as described in \cite{simon2024review}. 
Already for isotropic particles, the training of a functional fully applicable to a general, three-dimensional inhomogeneity, is currently a yet unsolved challenge for the machine learning approaches to DFT.  Here, the use of theory-informed \textit{ans\"atze} coming from FMT (as used here) or other weighted--density forms (see e.g. \cite{cats2021}) with a subsequent fitting of weight functions appears to be viable alternative to deep network representations \cite{sammuller2023neural,sammuller2024neural_ii,dijkman2024learning,sammuller2024pair, sammuller2024neural2}.      
A similar approach could be useful in MDFT to design an ML-enhanced bridge functional which goes beyond the simple hard sphere bridge functional. 
If one wants to include the angular dependency of the density profiles into an informed ML model, the use of orientational expansions appears to be mandatory in order to reduce degrees of freedom. In this respect, our results constitute an encouraging example of the rapid convergence of the orientational expansion for a strongly anisotropic model fluid, which, however, comes at a higher price of numerical complexity. 

\section{Supplementary Material}
As supplementary material we present additional data for the correlation
functions (pair correlation function, direct correlation
function, bridge function) at various state points which
further illustrates the discussion from the main text.

\begin{acknowledgments}
We gratefully acknowledge funding by the Deutsche Forschungsgemeinschaft (DFG, German Research Foundation) under Germany’s Excellence Strategy EXC no. 2064/
1, Project no. 390727645.
LB and DB acknowledge support from the Agence Nationale de la Recherche, projet ANR BRIDGE AAP CE29.
A.\ Simon is grateful for the hospitality during a research stay in Paris where part of this work was completed. 
\end{acknowledgments}

\nocite{*}
\bibliography{main} 
%\pagebreak
%\input{si.tex}
\end{document}

\title{Supplementary Information}
\maketitle
%section{

Here we present additional data for the correlation functions (pair correlation function, direct correlation function, bridge function) at various state points which further illustrates the discussionfrom the main text.

\section{Radial distribution functions}
\begin{figure*}
    \centering
    \begin{subfigure}[t]{\textwidth}
  \centering
    \includegraphics[width=0.8\linewidth]{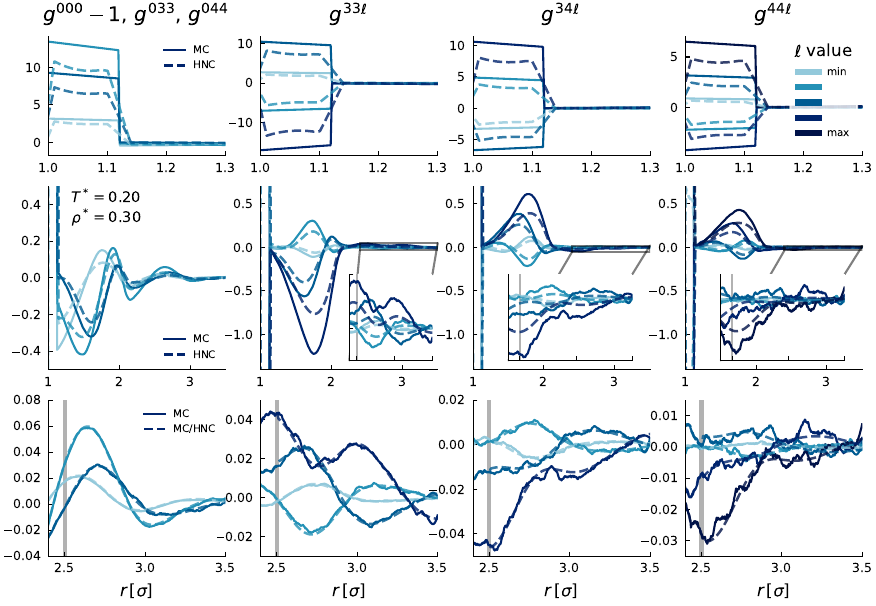}%
    \end{subfigure}
    \newline
    \begin{subfigure}[t]{\textwidth}
  \centering
    \includegraphics[width=0.8\linewidth]{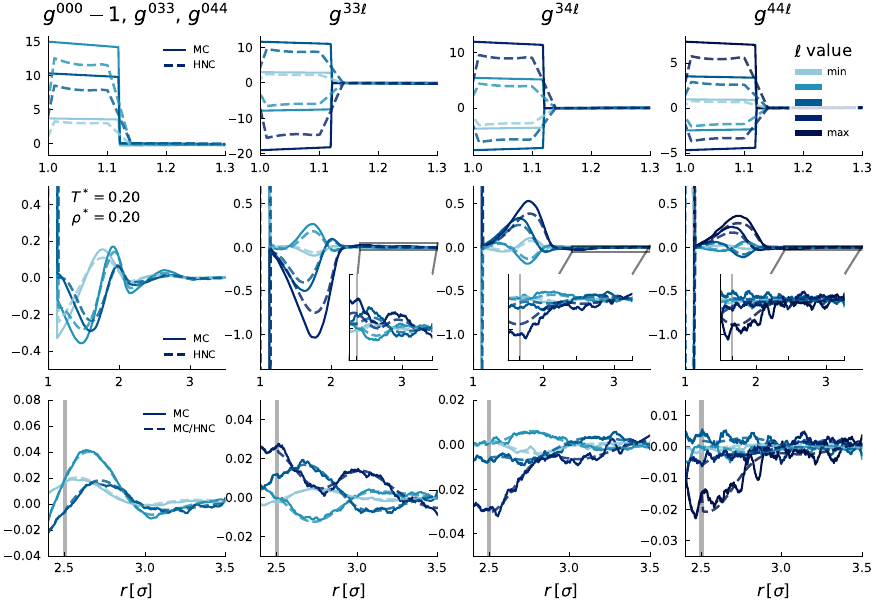}   
  \end{subfigure}
  \caption{The full RDF at the two state points $\rho^*=0.3, 0.2$ and $T^*=0.2$. The upper two
    rows show the MC results together with the HNC integral equation solutions. The lower row
    shows the results of the combined MC/HNC integral equation closure after the cut-off radius.}
  \label{fig:gfull}%
\end{figure*}

The radial distribution functions (RDF) for the patchy particle
fluid are shown are shown in \Cref{fig:gfull} for at different state points and
for all non-vanishing moments up to the cut-off $l_\mathrm{max} = 4$.
In constrast to the main text, no  rescaling w.r.t.\ the homogenouse RDF $g^{000}$
has been performed for easier comparison. The first row shows the HNC solution together with the
MC results in the bonding region, while the second row shows the
mid-range region.
Finally, in the third row, the solutions of integral equation with the MC/HNC  
closure are shown (also omitted in the main text). As by construction the profiles are equal to the MC curves
for $r \leq 2.5$ (cut-off shown as gray bar), we show only the relevant region further out.
\section{Bridge functions}
\begin{figure*}
\centering
\begin{subfigure}[t]{.48\textwidth}
  \centering
  \includegraphics[width=\linewidth]{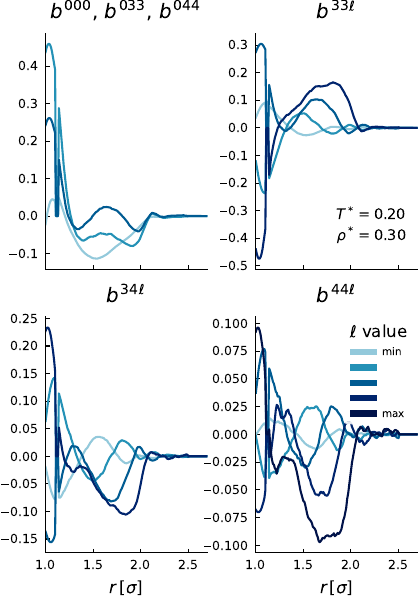}%
\end{subfigure}
\hfill
\begin{subfigure}[t]{.48\textwidth}
  \centering
  \includegraphics[width=\linewidth]{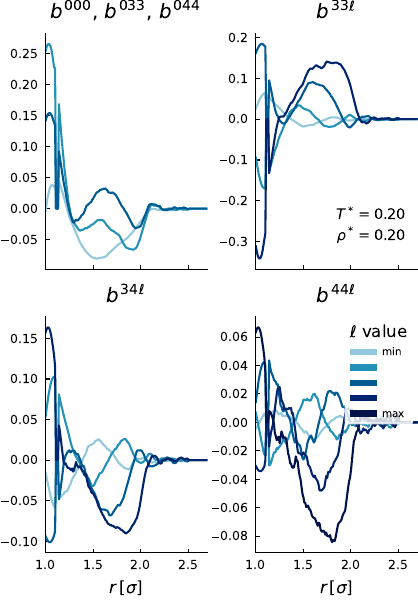}%
\end{subfigure}
\caption{Bridge function extracted from simulations for two state points $T^* = 0.20$ and $\rho^* = 0.2, 0.3$.}
  \label{fig:bridgefull}
\end{figure*}

As described in the main text we can extract from the MC/HNC solutions the bridge function.
The HNC condition, i.e. $b=0$, is imposed from $r = 2.5$ on, which marks the point where
the bridge functions must definitely vanish. In \Cref{fig:bridgefull} we show the computed bridge functions
up to the cut-off $l_\mathrm{max}=4$. One can see that (up to statistical noise) all
curves tend toward zero, well before the cut-off at $r=2.5$. As explained in the
main text, this validates the choice of the cut-off.
\section{Direct correlation function}
\begin{figure*}
\centering
\begin{subfigure}{\textwidth}
\centering
\includegraphics[width=0.8\columnwidth]{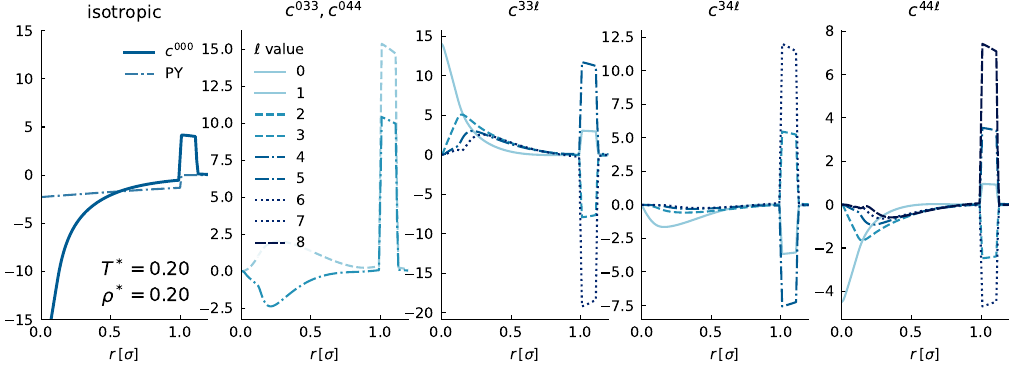}%
\end{subfigure}
\newline
\begin{subfigure}{\textwidth}
\centering
\includegraphics[width=0.8\columnwidth]{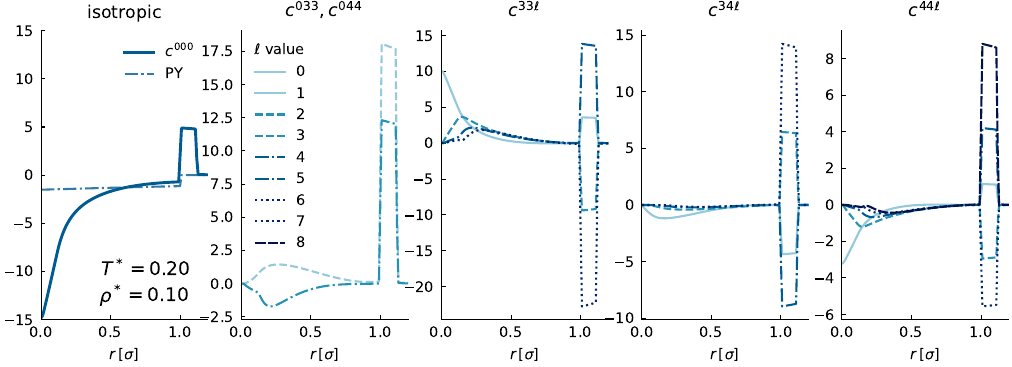}
\end{subfigure}
\newline
\begin{subfigure}{\textwidth}
\centering
\includegraphics[width=0.8\columnwidth]{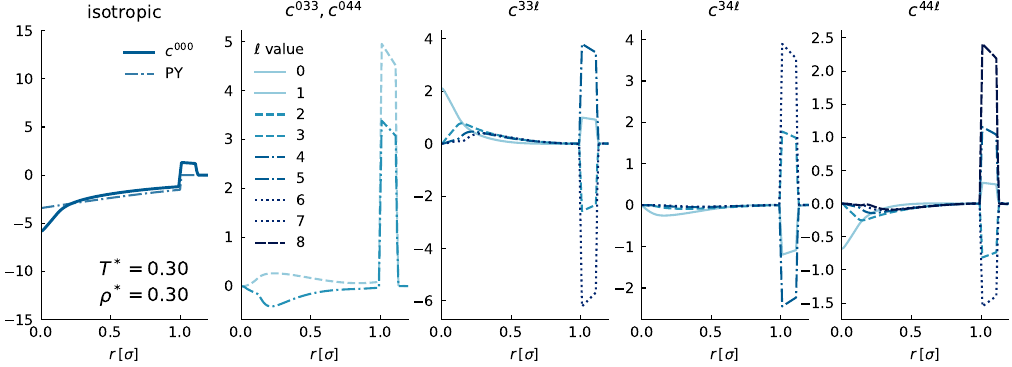}
\end{subfigure}
\newline
\begin{subfigure}{\textwidth}
\centering
\includegraphics[width=0.8\columnwidth]{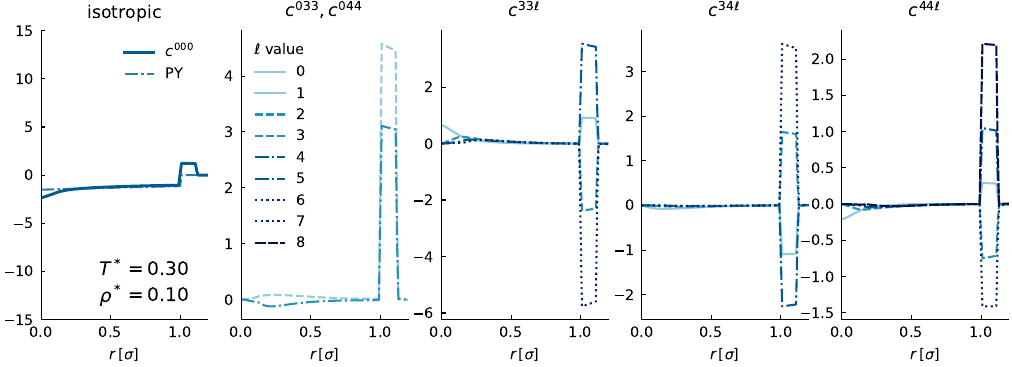}
\end{subfigure}
  \caption{Moments of the DCF $c(r)$ for multiple state points. The angular averaged (isotropic) moment $c^{000}$ is shown in comparison to the PY DCF.}
  \label{fig:cfull}
\end{figure*}

Finally we show the complete expansion (again up to $l_\mathrm{max}=4$) of the direct correlation function extracted
from MC data of the KF fluid in \Cref{fig:cfull}. In addition, the IE solution (using the Percus-Yevick closure) of the DCF for the hard-sphere fluid is shown to illustrate the influence of the attractive interactions on the DCF. 
